# Frequency Stable Microwave Sapphire Oscillators

Eugene N. Ivanov and Michael E. Tobar

***Abstract*** **- We show that state-of-the-art phase noise and high frequency stability could be simultaneously achieved in a microwave oscillator based on the sapphire-loaded cavity resonator. The 9 GHz sapphire oscillator was constructed with the SSB phase noise close to -170 dBc/Hz at an offset frequency of 10 kHz and fractional frequency instability less than 2 $10^{-13}$ for integration times from 5 to 50 s. In this work, we focus on the technique for phase-referencing the microwave sapphire oscillator to a stable radio-frequency source. We also discuss the suppression of the fast phase fluctuations of the microwave signal due to its transmission through the high-Q resonator.**

***Index Terms*** **- microwave sapphire resonator, phase noise measurements, phase-locked loop, Allan deviation of fractional frequency fluctuations**

## I. INTRODUCTION

THE progress in laser frequency stabilization and optical frequency synthesis resulted in impressive advances in generating spectrally-pure microwave signals [1-6]. Recently, a team of NIST scientists showed that exceptional long-term frequency stability of the cold-atom optical clock could be transferred to the microwave domain via the optical frequency division technique [7]. The optical frequency synthesis was also applied to generate low-phase noise microwaves [8-11]. For example, [11] describes an optical synthesis of the 12 GHz signal with Single-Sideband (SSB) phase noise close to -173 dBc/Hz at Fourier frequency $F$ = 10 kHz. On the other hand, the phase noise comparable to [11] was reported for microwave sapphire oscillators with interferometric signal processing [12]. Such oscillators, until now, have suffered from a serious drawback – poor frequency stability caused by the strong dependence of sapphire dielectric permittivity on temperature. In this work, we describe the noise properties of the microwave sapphire oscillator with fractional frequency stability approaching that of the best Oven-Controlled Crystal Oscillators (OCXO).

## II. RESULTS AND DISCUSSION

Fig. 1 shows the schematic diagram of the microwave loop oscillator with interferometric signal processing. It is based on the Sapphire Loaded Cavity (SLC) resonator excited in whispering gallery mode at a frequency close to 9 GHz. A detailed description of such an oscillator was given in [12].

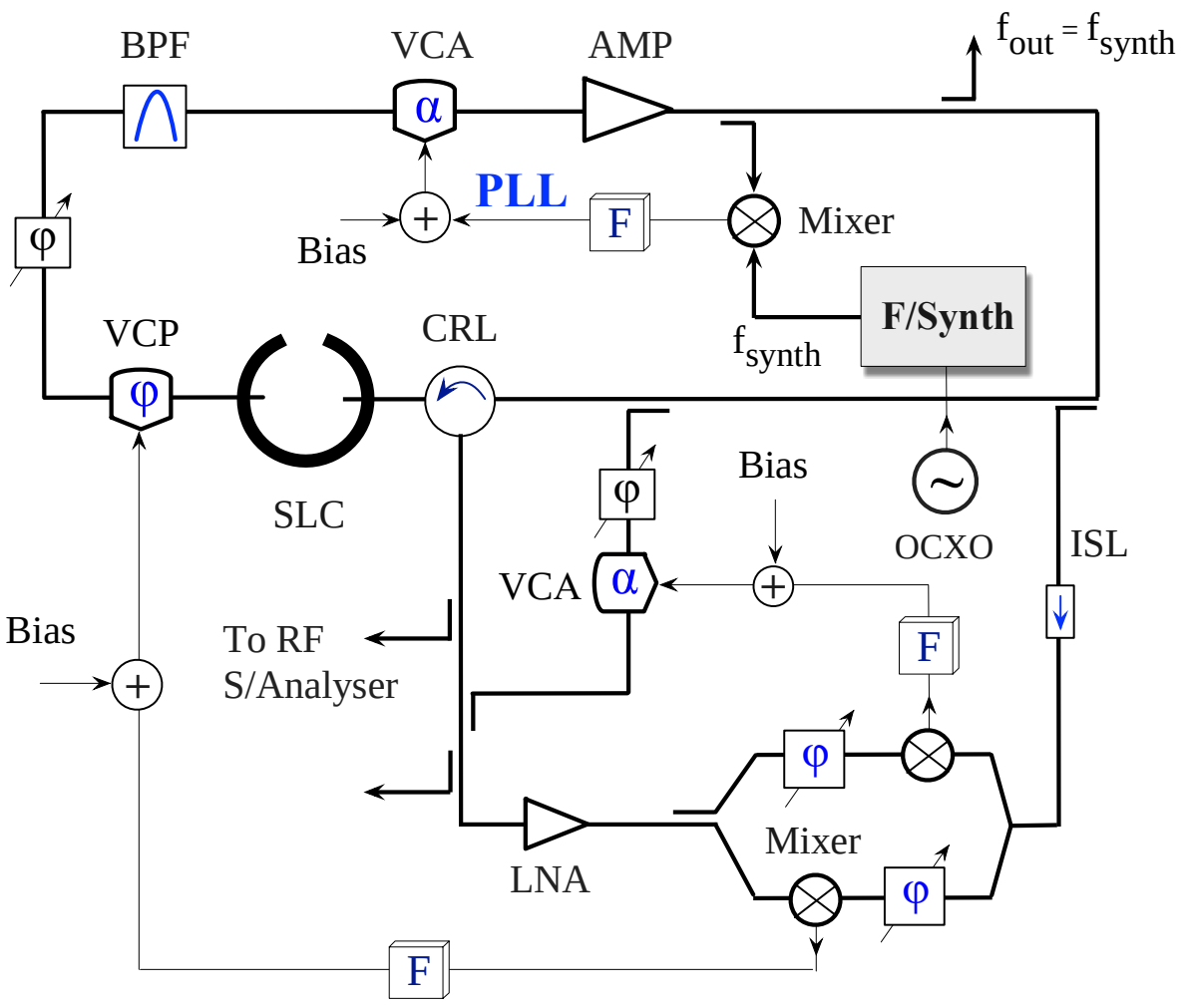

Fig. 1. Schematic diagram of microwave oscillator: BPF – band-pass filter, VCA – voltage-controlled attenuator, OCXO- oven-controlled crystal oscillator, CRL – circulator, VCP – voltage-controlled phase shifter, LNA – low-noise microwave amplifier, SLC – sapphire-loaded cavity resonator, ISL – isolator, F/Synth is the frequency synthesizer, PLL – phase-locked loop.

Also shown in Fig.1 is the Phase-locked Loop (PLL) referencing the microwave oscillator to the OCXO. A Voltage-Controlled Attenuator (VCA) in front of the loop amplifier (AMP) acts as the PLL actuator. It allows one to tune the oscillator frequency by controlling microwave power dissipated in the sapphire resonator. Below we consider the key points related to the phase locking of the microwave sapphire oscillator to the OCXO.

As the first step in the PLL design, we measured the frequency tuning coefficient of the sapphire oscillator $df_{osc}/du$ as a function of Fourier frequency. These measurements involved: (i) phase-locking of an external microwave frequency synthesizer (Keysight 8257D) to the sapphire oscillator; and (ii) measurements of the complex transfer function $\mathcal{H}$ between the VCA control port and frequency synthesizer FM-modulation port (see Fig. 2).

At Fourier frequencies within the bandwidth of the PLL in Fig. 2, the transfer function $\mathcal{H}$ is reduced to the ratio of the $df_{osc}/du$ to the tuning coefficient of the frequency synthesizer. Fig. 3 shows the magnitude of the $df_{osc}/du$ inferred from the measured transfer function $\mathcal{H}$. The inclined part of the $df_{osc}/du$ is due to the power-to-frequency conversion in the SLC

[1] Eugene Ivanov, Quantum Technologies and Dark Matter Labs, Department of Physics, The University of Western Australia, 35 Stirling Highway, Crawley, 6009 (eugene.ivanov@uwa.edu.au)
This work is supported by the Australian Research Council (grants CE170100009 and CE200100008)

Michael Tobar, Quantum Technologies and Dark Matter Labs, Department of Physics, The University of Western Australia, 35 Stirling Highway, Crawley, 6009 (michael.tobar@uwa.edu.au)

resonator; the horizontal pedestal is due to the VCA residual voltage-to-phase conversion. From the viewpoint of feedback control, the VCA-based frequency actuator corresponds to a "leaking" integrator with a time constant close to 50 s.

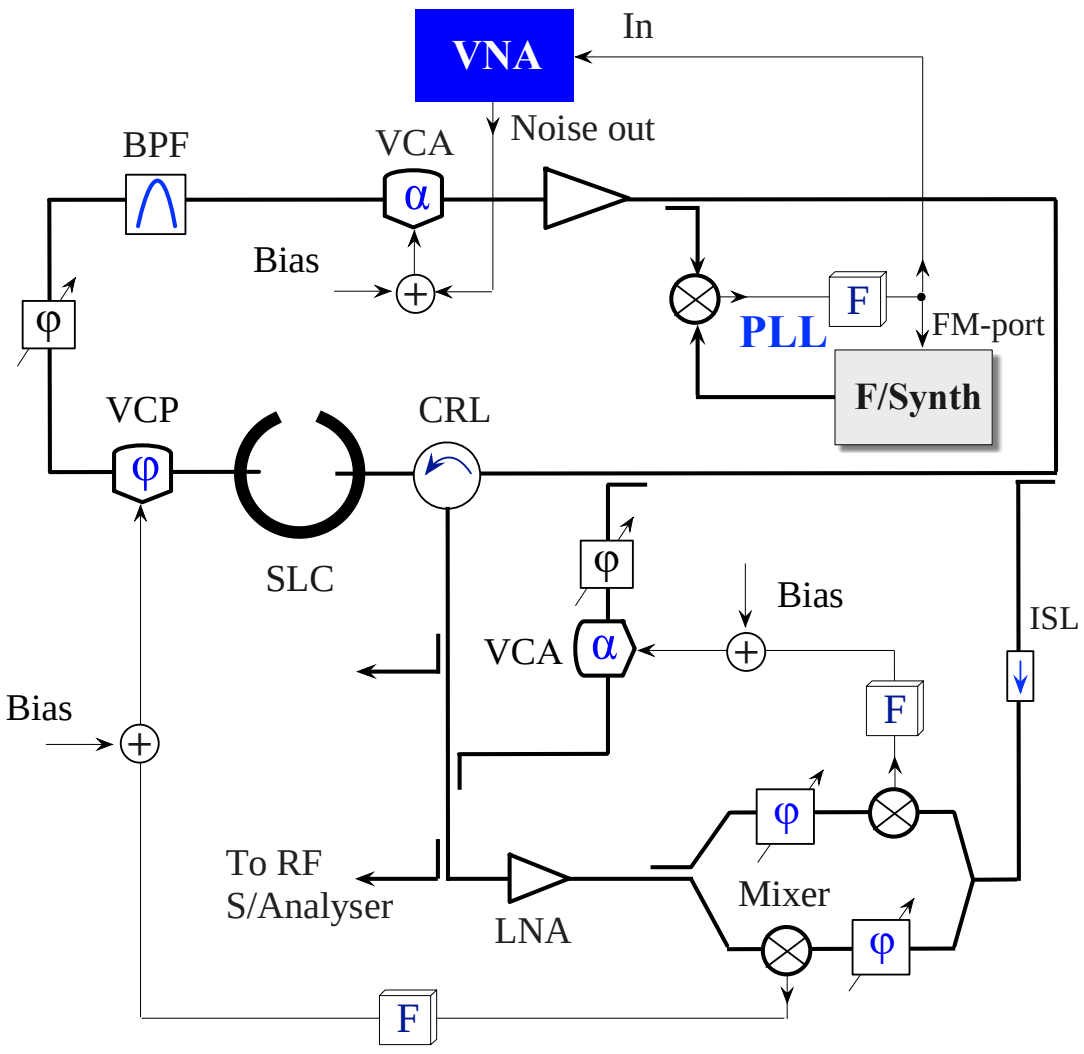

Fig. 2. Measurements of the frequency tuning coefficient of the sapphire oscillator: VNA is the vector network analyzer

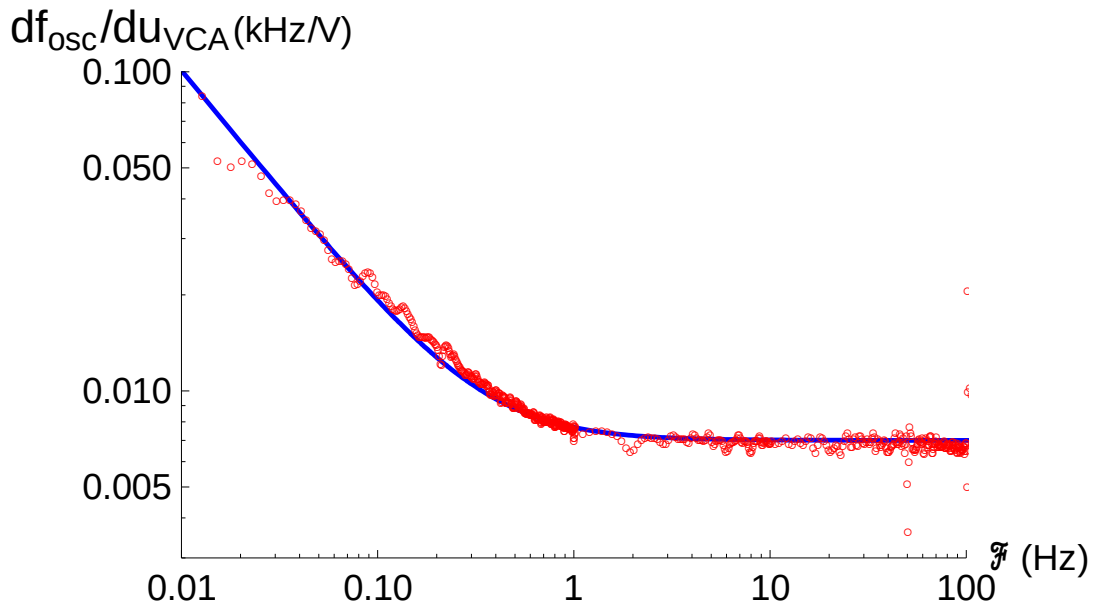

Fig. 3. Magnitude of voltage-to-frequency conversion

Having characterized the PLL actuator, we built an active filter for phase referencing the sapphire oscillator to OCXO (Fig. 4). The first stage of the filter is a quasi-differentiator with corner frequencies of 5 and 200 mHz. It ensures a stable operation of the PLL by introducing a phase advance into the feedback loop to compensate for the phase lag associated with the power-to-frequency conversion in the SLC resonator. The second stage of the filter is a "leaking" integrator with a time constant of approximately 1.5 s. It is needed to avoid PLL cycle slips due to the random walk of the oscillator frequency. The third stage of the filter is a summing device that combines the VCA bias voltage with the correction signal.

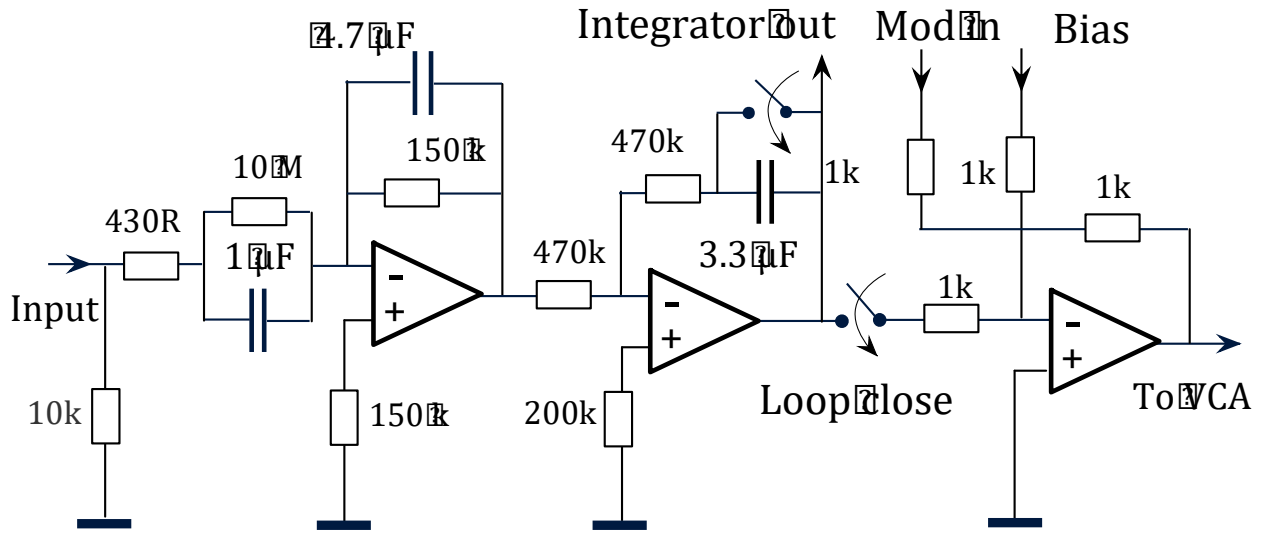

Fig. 4. PLL filter

A correctly designed PLL, apart from allowing the reliable phase tracking of the reference oscillator, must also be efficient in suppressing its phase fluctuations outside the loop bandwidth. Both these conditions were met in our case. First, we measured the SSB phase noise of the synthesizer $\mathcal{L}_{\varphi}^{synth}$ and sapphire oscillator $\mathcal{L}_{\varphi}^{SLCO}$. They proved to be almost equal at F = 1 Hz, whereas, at F = 10 Hz, the SSB phase noise of the sapphire oscillator was 20 dB less than that of the synthesizer.

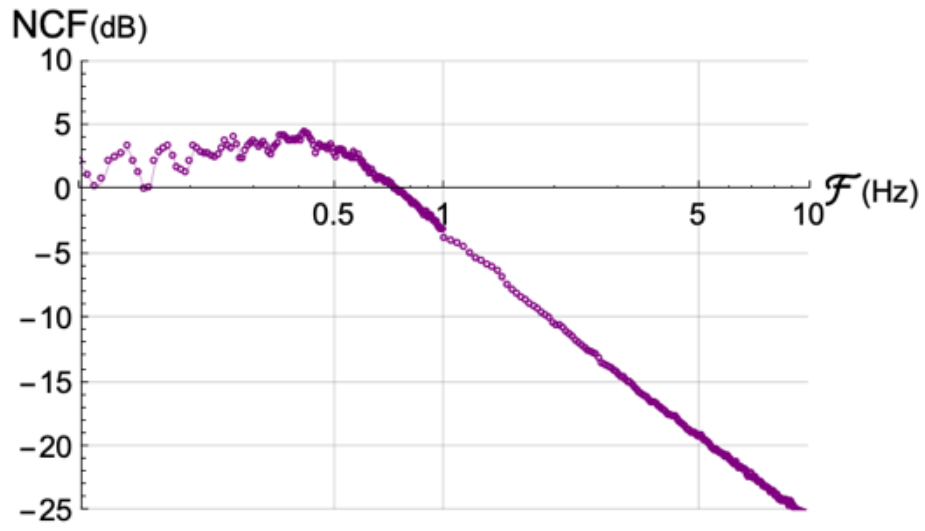

Fig. 5. PLL noise conversion factor

Next, we measured the amplitude transfer function between "Mod in" and "Integrator out" ports of the PLL filter in Fig. 4. Such a function characterizes the conversion of synthesizer phase noise into that of the sapphire oscillator and can be termed the PLL Noise Conversion Factor (NCF). Fig. 5 shows the NCF as a function of Fourier frequency. At F> 1Hz, the NCF drops rapidly with frequency and is sufficiently small to prevent any excess phase noise in the spectrum of the sapphire oscillator due to its phase locking.

Two microwave oscillators were constructed, each featuring the SLC resonator with an intrinsic Q-factor close to 200000. Fig. 6 shows a photo of one of the sapphire oscillators. The frequency separation between the oscillators can be adjusted within a few MHz by changing the temperature of one of the sapphire crystals.

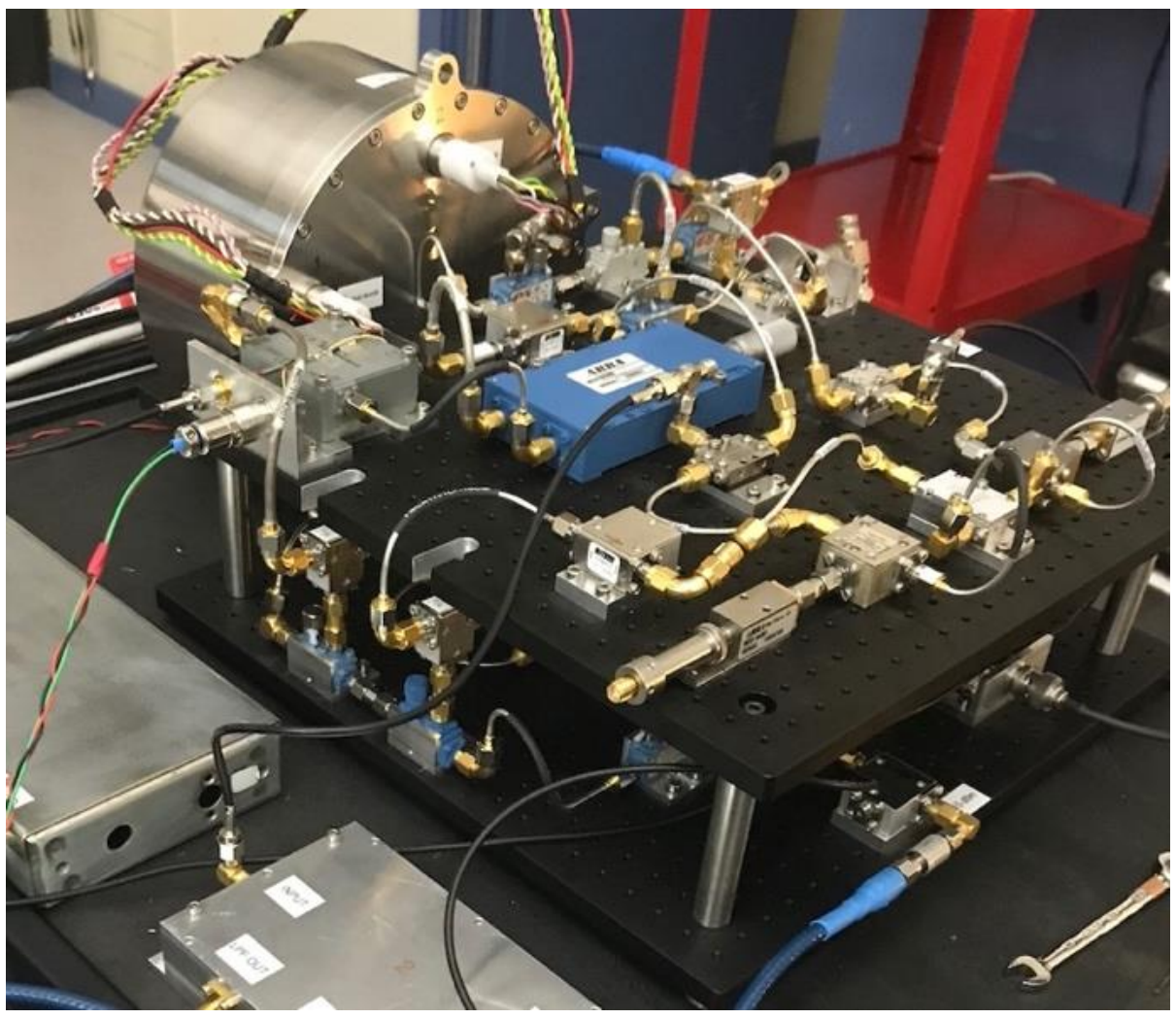

Fig. 6. A laboratory prototype of the low phase noise 9 GHz oscillator: the sapphire resonator is inside the cylindrical vacuum can; the top plate houses microwave components that form an interferometric frequency discriminator; the bottom plate accommodates oscillations sustaining stage.

Fig. 7 shows the phase noise measurements setup based on the four-channel digital phase detector 35100A (Microchip

Technology). The four-channel readout permits more accurate phase noise measurements than its two-channel counterpart due to the reduced uncertainties associated with the reference oscillators and mixers.

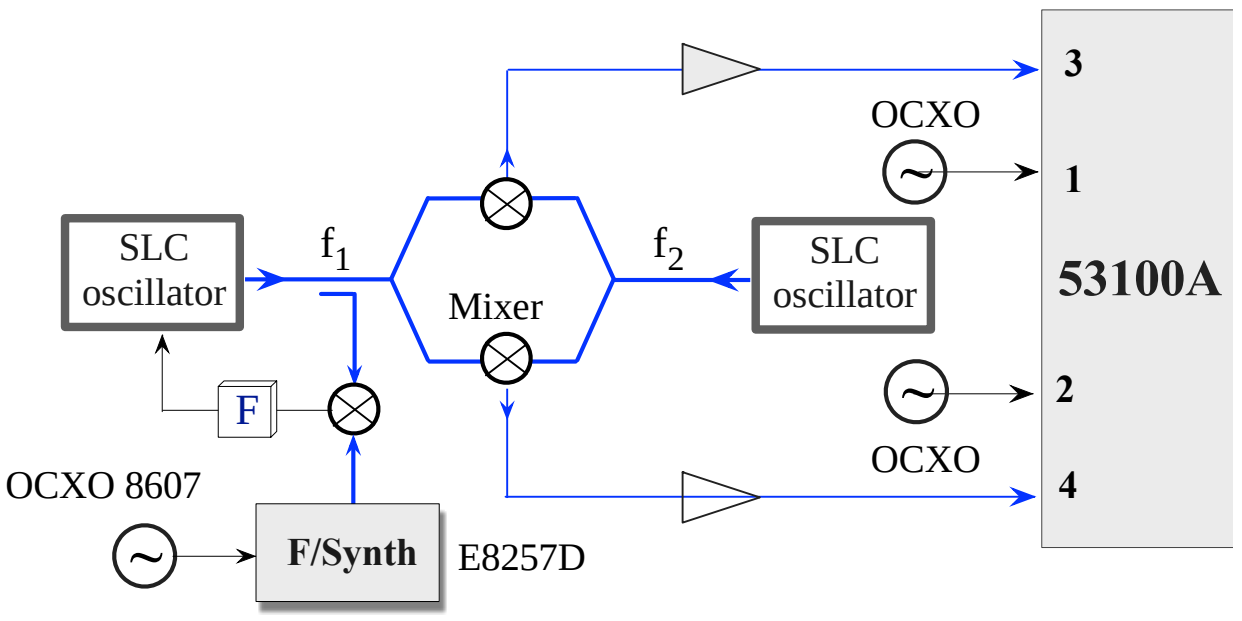

Fig. 7. Phase noise measurements system: the 53100A is a digital phase noise detector; frequency of the beat note between two microwave oscillators is close to 2 MHz

Fig. 8 shows the phase noise spectra of the frequency synthesizer (top trace) and sapphire oscillator (bottom trace). The bottom trace is, in fact, two almost indistinguishable traces corresponding to the open and closed phase-locked loop, which confirms our early estimate based on the measurements of the PLL NCF.

At Fourier frequencies from 20 to 100 kHz, oscillator phase noise $\mathcal{L}_{\varphi}^{SLCO} < -170\ dBc/Hz$, which is comparable with the best noise levels reported in [12]. On the other hand, at F < 3 kHz, the phase noise of the sapphire oscillator is not as low as in [12]. This is because of the residual phase noise of the VCA placed inside the interferometer.

In the next stage of experiments, we studied how efficiently one can suppress the oscillator phase noise by extracting the output signal from the resonator transmission port. In [13], it was shown that the resonator in transmission acts as a band-pass filter suppressing both phase and amplitude fluctuations of the output signal outside its bandwidth, provided that the resonator transmission peak has sufficiently high contrast. Both SLC resonators in our experiments displayed slightly distorted Lorentzian transmission profiles with a 20…25 dB contrast.

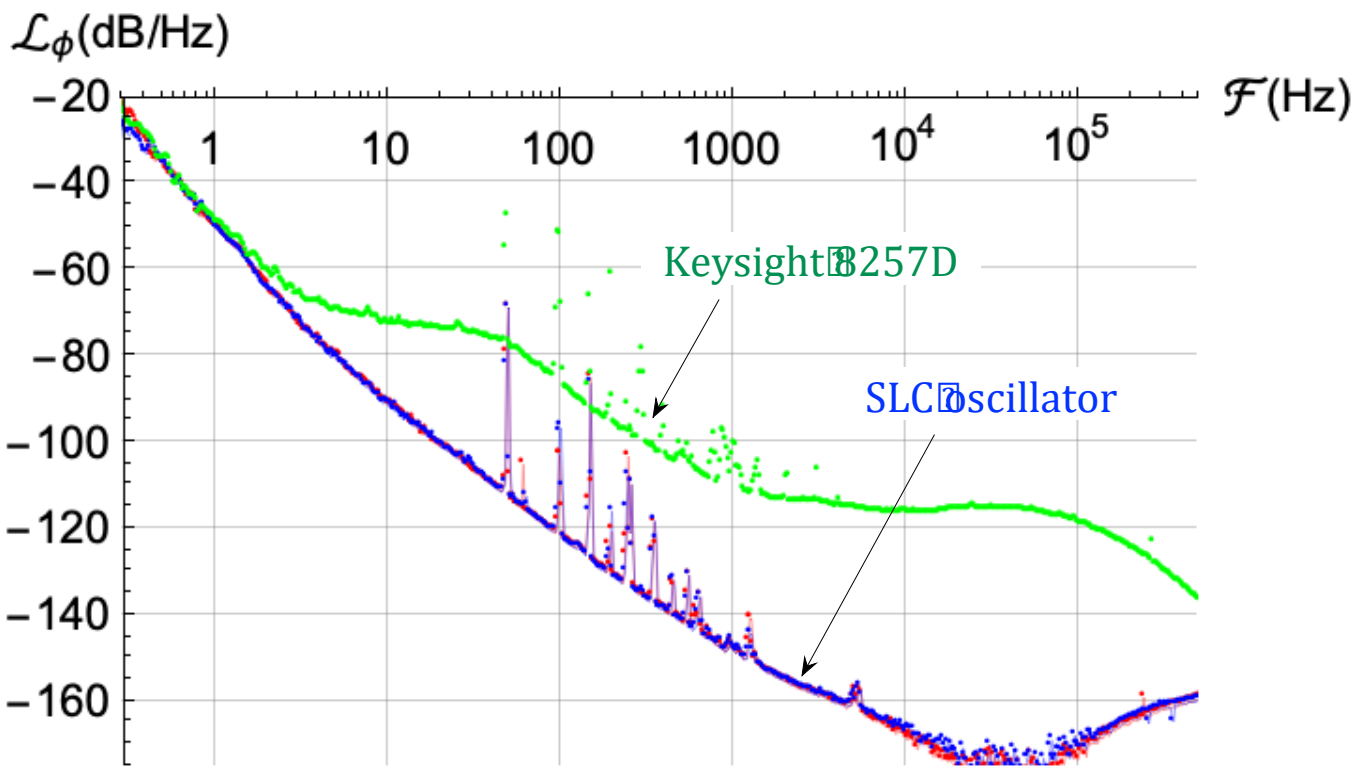

Fig. 8. Phase noise spectra at 9 GHz

We introduced a Wilkinson power divider into each oscillator to divert half of the transmitted power (3…5 dBm) to the noise measurements system. As expected, at F < 30 kHz, the phase noise spectra of the incident and transmitted signals closely followed each other, whereas, at F > 50 kHz, i.e., outside the SLC resonator bandwidth, they diverged. The phase noise intensity of the incident signal was rising with Fourier frequency (bottom trace in Fig. 8). In contrast, the phase noise intensity of the transmitted signal at F>50 kHz was close to -175 dBc/Hz, which was the spectral resolution limit of the 53100A.

Finally, we evaluated the short-term frequency stability of the phase-locked sapphire oscillator. We used Keysight 8257D phase-locked to 8607 OCXO from Oscilloquartz as a frequency reference. The beat note frequency was 300 kHz, and counter gate time was 1 s. The data were collected for 4 hours. Fig. 9 shows Allan deviation of fractional frequency fluctuations of the SLC oscillator as a function of integration time. For integration times 5…50 s, the fractional frequency instability of the sapphire oscillator is slightly below 2 $10^{-13}$, which is consistent with the OCXO manufacturer's specifications. It should be noted that Allan deviation was computed using the frequency counter "raw" data; no attempt to exclude the frequency drift was made.

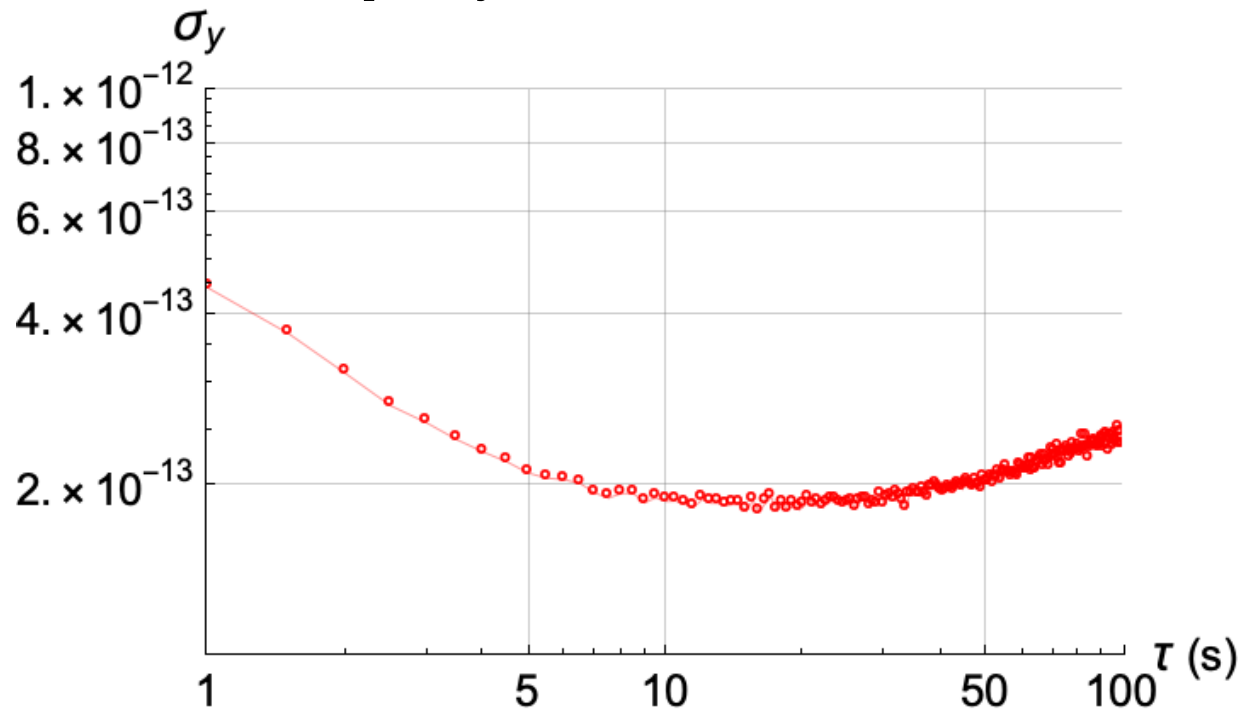

Fig. 9. SLC oscillator fractional frequency stability

## III. CONCLUSION

Our experiments showed that:

- The microwave sapphire oscillators can rival the best radio-frequency signal sources in terms of their fractional frequency stability;
- The high-frequency stability of the microwave sapphire oscillators can be achieved without any noticeable degradation of their phase noise performance;
- The phase noise of the microwave sapphire oscillators can be reduced at high Fourier frequencies by extracting the output signal from the resonator transmission port, provided that the contrast of the resonator transmission peak is sufficiently high (>20 dB).

Compared to optoelectronic oscillators based on femtosecond laser technology, microwave sapphire oscillators are characterized by small footprints and low g-sensitivity. These features make them well-suited for various mobile applications [14, 15].

## ACKNOWLEDGEMENT

We acknowledge the valuable technical assistance of the South Australian company QuantX Labs, which provided two temperature-stabilized sapphire resonators for the project.